\documentclass[twocolumn,aps,prl]{revtex4-1}

	\usepackage[usenames,dvipsnames]{xcolor}
	\usepackage{amsmath}
	\usepackage{amsfonts}
	\usepackage{amssymb}
	\usepackage[colorlinks=true,citecolor=blue,linkcolor=blue,urlcolor=blue]{hyperref}
	\usepackage{graphicx}
	\usepackage{bbold}					
	\usepackage[makeroom]{cancel}		
	\usepackage{multirow}				
	\usepackage[normalem]{ulem}        
	\usepackage{array}
	\usepackage{pdfpages}
    \usepackage{float}
    \usepackage{upgreek}

	\makeatletter
	\AtBeginDocument{\let\LS@rot\@undefined}
	\makeatother
	
	\newcolumntype{x}[1]{>{\centering\let\newline\\\arraybackslash\hspace{0pt}}p{#1}}

	\DeclareMathAlphabet{\mathbbold}{U}{bbold}{m}{n}

	\newcounter{subeqn} %
	\renewcommand{\Re}{\operatorname{Re}}

	\makeatletter
	\@addtoreset{subeqn}{equation}
	\makeatother

\definecolor{TB}{rgb}{1,0.5,0}

\def\beq{\begin{equation}}
\def\eeq{\end{equation}}
\def\bald{\begin{aligned}}
\def\eald{\end{aligned}}
\def\bea{\begin{eqnarray}}
\def\eea{\end{eqnarray}}

\def\Tr{\mathrm{Tr}}

\def\Eq#1{Eq.~(\ref{#1})}
\def\Fig#1{Fig.~\ref{#1}}

\begin{document}

\title{Diagnosis of mixed-state topological phases in strongly correlated systems via disorder parameters}

\author{Shao-Hang Shi$^{1,2}$}
\author{Xiao-Qi Sun$^{3,1}$}
\email{xiaoqi.sun@mpq.mpg.de}
\author{Zi-Xiang Li$^{1,2}$}
\email{zixiangli@iphy.ac.cn}

\affiliation{$^1$Beijing National Laboratory for Condensed Matter Physics $\&$ Institute of Physics, Chinese Academy of Sciences, Beijing 100190, China}
\affiliation{$^2$University of Chinese Academy of Sciences, Beijing 100049, China}
\affiliation{$^3$Max Planck Institute of Quantum Optics, Hans-Kopfermann-Straße 1, D-85748 Garching, Germany}

\date{\today}

\begin{abstract}
Characterizing topological phases for strongly interacting fermions in the mixed-state regime remains a major challenge. Here we introduce a general and numerically efficient framework to diagnose mixed-state topological phases in strongly interacting systems via the disorder parameter (DP) of the U(1) charge operator.  Specifically, from the finite-size scaling of the second derivative of the DP generating function, we introduce the  topological scaling indicator, which exhibits a characteristic linear scaling with the system's linear dimension for topological phases, a signature that vanishes upon transition into a topologically trivial phase. Crucially, we develop an efficient determinant Quantum Monte Carlo algorithm that facilitates the evaluation of this indicator in interacting systems. We apply our approach to two paradigmatic models: for the Kane-Mele-Hubbard model, we successfully map the interaction-driven transition from a quantum spin Hall insulator to a trivial Mott insulator. Furthermore, our method circumvents the limitations imposed by the severe sign problem in the Haldane-Hubbard model, enabling robust identification of the quantum anomalous Hall phase at accessible temperatures.  This work provides a powerful and accessible tool for the numerical exploration of topological phenomena in interacting mixed states, opening a pathway to study systems previously inaccessible due to computational obstacles.
\end{abstract}

\maketitle
{\em Introduction.}---Inevitable coupling to the environment turns pure states into mixed states, posing challenges to define robust signatures identifying phases in the presence of noise. While topological phases for mixed states have been discussed using a variety of theoretical approaches (see e.g.,~\cite{coser2019quantum,degroot2022quantum,Wang2023PRX,Sang2024,Wang2025PRX,zhongwang2025prx,Cheng2025PRXQuantum,sohal2025prxquantum,sang2025mixed,Yang2025PRX,Lee2025Quantum,Fan2024PRXQuantum,Pal2024PRXQuantum,Oshikawa2024PRB,Ellison2025PRL,Ashida2024PRB}), recent proposals have introduced non-local quantities~\cite{viyuela2014prl1d,budich2015prb, bardyn2018prx,altland2021prx,huang2022prb,huang2025mixed,mao2024rpp,huang2025prlnew}, such as the ensemble geometric phase~\cite{bardyn2018prx, huang2022prb}, as characteristic features extending the notion of topological phases to mixed states, with sharp signatures detectable on emerging experimental platforms. These experimental platforms, e.g., trapped ions~\cite{blatt2012nature,monroe2021rmp,noel2022np}, Rydberg atoms~\cite{semeghini2021science}, and superconducting circuits~\cite{satzinger2021science}, enable measuring non-local observables, such as the full-counting statistics~\cite{gross2017science}, beyond conventional solid-state devices~\cite{blatt2012nature,monroe2021rmp,noel2022np,gross2017science, semeghini2021science, ebadi2021nature, satzinger2021science,iqbal2024nature, acharya2024nature,chen2025np}. 

Despite recent progress in characterizing mixed-state topological phases, a comprehensive understanding of mixed-state topology in strongly correlated systems remains elusive. For free-fermion cases, i.e., mixed states described by fermionic Gaussian states, nonlocal topological order parameters have been defined to characterize mixed-state symmetry-protected topological (SPT) phases~\cite{viyuela2014prl1d,budich2015prb, bardyn2018prx,altland2021prx,huang2022prb,huang2025mixed,mao2024rpp}. However, extending these concepts to strongly interacting fermionic systems has proven difficult, with progress largely confined to one-dimensional models~\cite{huang2025prl}, leaving the topological physics of interacting mixed states in higher dimensions a crucial and largely unexplored frontier. As the many-body Hilbert space dimension grows, especially at higher dimensions, the fundamental problem lies in defining characteristic quantities that are efficient to probe both numerically and experimentally. 

In this Letter, we introduce an efficient and general approach to identify the topology of a density matrix $\hat{\rho}$ using the disorder parameter (DP)~\cite{Levin2020,Cheng2025PRB,Yao2024PRL,Zhang2024PRLFCS,Huang2025PRLQC,xu2025diagnosing,Wan2021PRB}, i.e., $Z(\theta) =  \Tr[\hat{\rho} e^{\text{i}\theta \hat{Q}}
]$ with $\hat{Q}$ being the charge operator, for strongly interacting systems of linear size $L$ with open boundary conditions (OBCs). The idea comes from the characterization of gapless topological boundary modes in the modular Hamiltonian for the free-fermion case, where the generating function $F(\theta)=L^{-d}\ln|Z(\theta)|$, with $d$ being the spatial dimension, exhibits a singularity at $\theta=\pi$. While $F(\theta)$ is inefficient to compute and measure since $Z(\theta)$ is exponentially small in $L$ around $\theta=\pi$, we propose the second derivative $F^{(2)}(\pi)$ as a characteristic signature of the topology of $\hat{\rho}$, revealing the gapless boundary modes in the modular Hamiltonian. Crucially, our proposal allows numerically exact quantum Monte Carlo (QMC) computations for higher dimensions with a speedup exponentially in $L$. In particular, this enables us to explore prototype examples such as Kane-Mele-Hubbard (KMH) model and Haldane-Hubbard (HH) model at finite temperatures, revealing qualitatively distinct features characterizing the topology of the density matrix for two dimensions: The finite-size scaling behavior of $F^{(2)}(\pi)$, from which we define the topological scaling indicator (TSI), distinguishes two phases with a linear scaling in the system's linear size for topological phases, and a nearly vanishing behavior for trivial phases. 

\begin{figure}[t]
    \centering
    \includegraphics[width=\columnwidth]{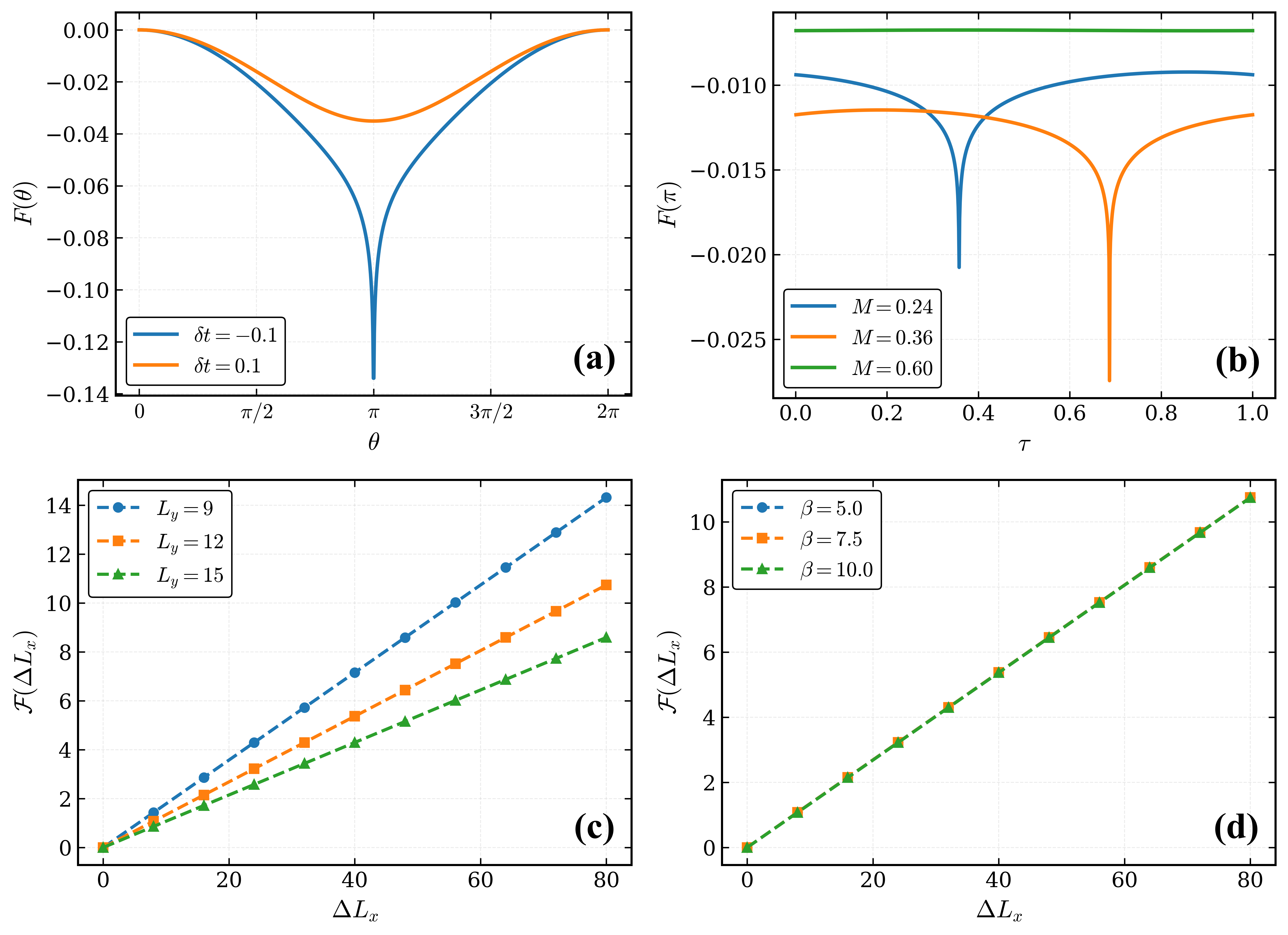}
    \caption{(a) $F(\theta)$ calculated from the 1D SSH model ($\beta=10.0$ and $L=128$). In the topological phase ($\delta t=-0.1$), a distinct cusp emerges at $\theta=\pi$, which is absent in the trivial phase ($\delta t=0.1$).
 (b) Response of $F(\pi)$ to a threaded magnetic flux $2\pi\tau$ in the Haldane model. A sharp cusp develops exclusively in the topological phase ($M < 3\sqrt{3}\lambda$). The parameters are \( t_1 = 1.0 \), \( \lambda = 0.1 \),  \(L_x= L_y = 36\) and \( \beta = 10.0 \).
(c,d) The topological scaling indicator $\mathcal{F}$ scales linearly with $\Delta L_x$ in the Haldane model ($M=0, \lambda=0.1$).
(c) The slope of $\mathcal{F}$ is proportional to $1/L_y$ at fixed $\beta=10$. 
(d) The results of $\mathcal{F}$ for fixed $L_y=12$ under different $\beta$. 
}
    \label{Fig1}
\end{figure}

{\em Non-interacting case.}---Let us start with the simple case of a Gibbs state for free fermions, where  $F(\theta)$ can be decomposed into the individual contribution from each single-particle eigenstate, i.e., $F(\theta) = \sum_{\epsilon} f_{\epsilon}(\theta)$ with $f_{\epsilon}(\theta)$ given by
\begin{equation}
\begin{aligned}
f_{\epsilon}(\theta)
= \frac{1}{L^d}
  \ln \sqrt{1-\frac{\sin^2(\theta/2)}{\cosh^2(\beta \epsilon/2)} }.
\end{aligned}
\end{equation}
Here, $\beta = 1/T$ is inverse temperature. This expression reveals a key feature: at $\theta=\pi$, the presence of a zero-energy mode ($\epsilon=0$) causes a logarithmic divergence, offering a sharp signature for boundary modes in topological systems. In particular, for 1D systems~\cite{mao2025probingurl}, the emergence of in-gap zero modes at the boundary is the defining feature for the topology of the Hamiltonian, leading to sharp cusp for $F(\theta)$ at $\pi$, with its height scaling as $\frac{1}{\beta}\frac{\mathrm{e}^{L}}{L}$.  
For illustration, we refer to the 1D Su-Schrieffer-Heeger (SSH) model~\cite{SSHmodel}:
\begin{equation}
\hat{H}_{\text{SSH}} = \sum_{i}^{L-1} \big[(t+\delta t)\hat{c}_{2i-1}^{\dagger}\hat{c}_{2i} + (t-\delta t)\hat{c}_{2i}^{\dagger}\hat{c}_{2i+1} + \text{h.c.}\big].
\end{equation}
As depicted in \Fig{Fig1}(a), the behavior of $F(\theta)$ clearly distinguishes the two phases: a pronounced cusp appears at $\theta=\pi$ in the topological regime ($\delta t < 0$), while the function remains smooth in the trivial regime ($\delta t > 0$).

We now extend our analysis to the 2D case by considering a cylindrical geometry with a periodic boundary conditions (PBC) along the $x$-direction and an OBC along the $y$-direction, which preserves the momentum $k_x$ as a good quantum number. A key distinction from the 1D case is that the edge states are no longer isolated zero-energy modes but form gapless bands that disperse with $k_x$. For a finite-size system, the quantization of $k_x$ opens a finite-size gap generically proportional to $\frac{2\pi}{L_x}$. 
Therefore, even for a topologically nontrivial system, the quantity $F(\pi)$ may remain finite, obscuring the characteristic cusp. We consider the Haldane model as an illustrative example~\cite{Haldane1988PRL}:
\begin{equation}
\begin{aligned}
\hat{H}_{\text{Hal}} = -t\sum_{\langle i,j \rangle}  \hat{c}_{i}^{\dagger}\hat{c}_{j} + + i\lambda \sum_{\langle\langle i,j \rangle\rangle, \sigma} \nu_{ij} \hat{c}^{\dagger}_{i} \hat{c}_{j} + M \sum_{i} \epsilon_i \hat{c}_{i}^{\dagger}\hat{c}_{i},
\label{Haldane}
\end{aligned}
\end{equation}
where  $\epsilon_i = \pm 1$ is a staggered on-site potential on the A/B sublattices. The imaginary next-nearest-neighbor hopping, proportional to $\lambda$, breaks time-reversal (TR) symmetry (with $\nu_{ij} = \pm 1$ for counter-clockwise/clockwise paths). The system is in a topologically nontrivial phase exhibiting the quantum anomalous Hall (QAH) effect when $M<3\sqrt{3} \lambda $. We consider the system on a $L_x \times L_y$ cylinder. To distinguish these phases, we thread a magnetic flux $2\pi\tau$ through the cylinder and monitor $F(\pi)$ as $\tau$ is varied adiabatically (shown in \Fig{Fig1}(b)). In the topological phase, the \( F(\pi) \) versus \(\tau\) curve exhibits a sharp cusp, whereas in the trivial phase, \( F(\pi) \) remains nearly unchanged.

\begin{figure}[t]
    \centering
    \includegraphics[width=\columnwidth]{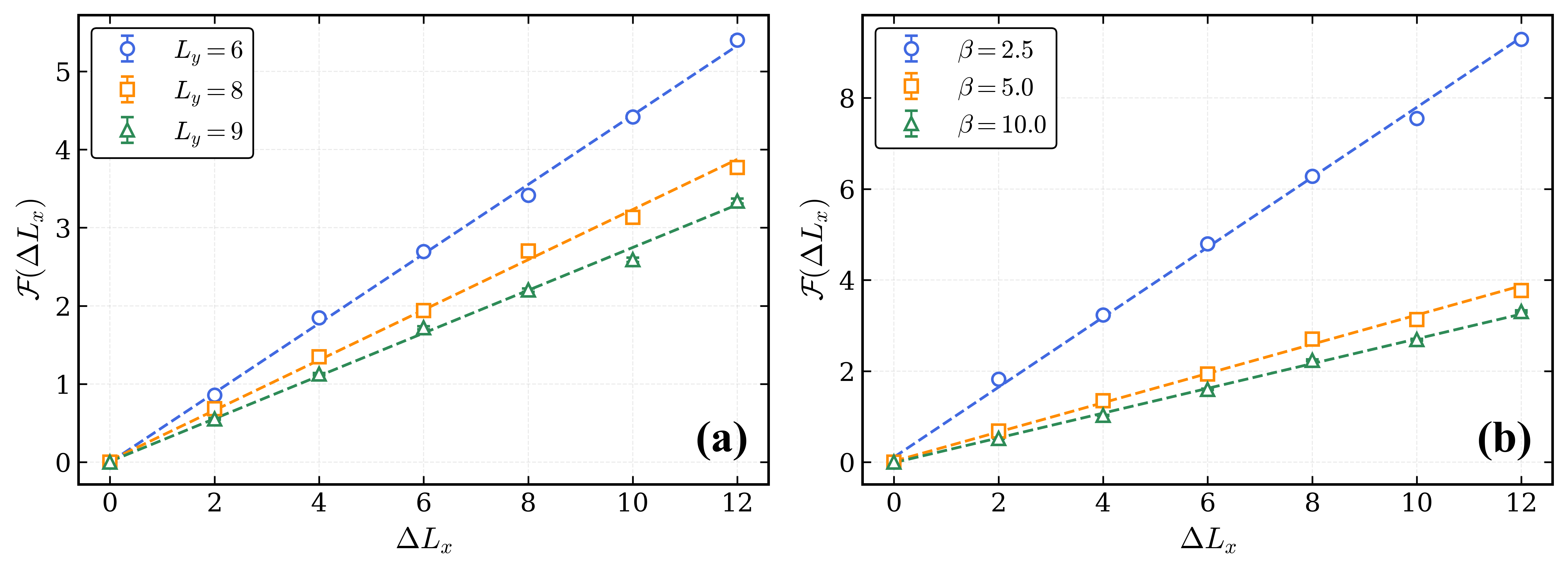}
    \caption{(a) Distinct linear scaling of $\mathcal{F}$ with $\Delta L_x$ for different $L_y$, calculated in the topological phase ($U = 2.5$) of KMH model at $\beta = 5.0$. (b) Linear scaling of $\mathcal{F}$ with $\Delta L_x$ for fixed $L_y = 8$ at different inverse temperatures $\beta$.  The reference system used in defining $\mathcal{F}$ has $L_x^0 = 7$.  }
    \label{Fig2}
\end{figure}

\begin{figure*}[htbp]
    \centering    \includegraphics[width=\textwidth]{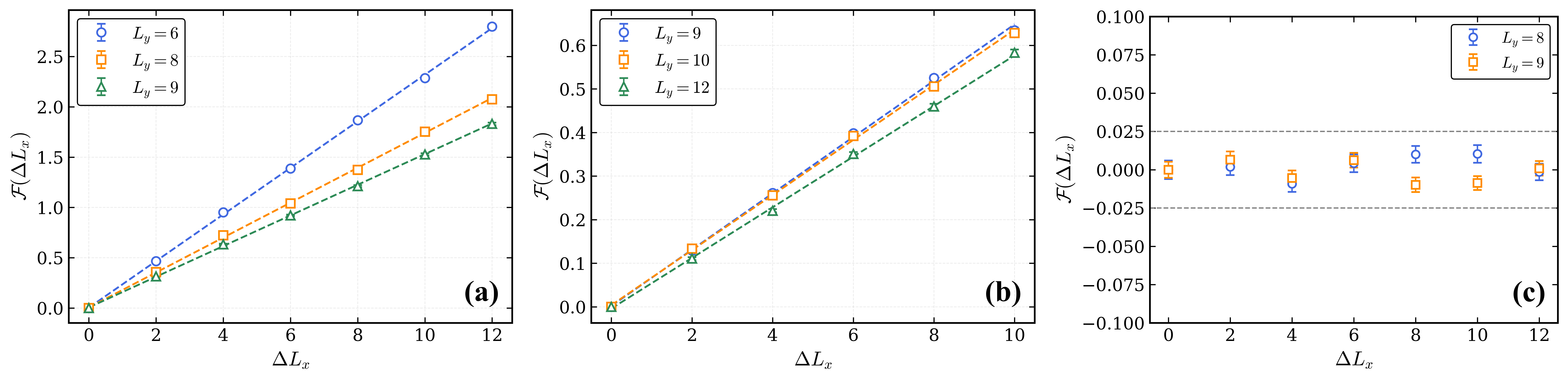}
    \caption{ The results of TSI $\mathcal{F}$ across the quantum phase transition in the KMH model. The inverse temperature is fixed at $\beta=8$.
(a) Deep in the topological phase ($U = 3.6$), $\mathcal{F}$ exhibits the expected linear relation with $\Delta L_x$. the reference system size is $L_x^0 = 7$.
(b) Approaching the phase transition ($U = 4.5$), the linear scaling remains robust, confirming the persistence of the topological phase even as correlations strengthen. The reference size is increased to $L_x^0 = 9$.
(c)In the trivial AFM insulating phase ($U=6.0$), the linear scaling breaks down entirely and the TSI is suppressed towards zero, correctly identifying the non-topological nature of the state.
    }
    \label{Fig3}
\end{figure*}

To characterize this singularity in finite systems quantitatively and in a more practical manner that avoids the necessity of tuning an external magnetic flux, we utilize the second derivative of $F(\theta)$ at $\theta=\pi$:
\begin{equation}
\begin{aligned}
F^{(2)}(\theta)\rvert_{\theta=\pi} = \sum_{\epsilon} f^{(2)}_{\epsilon}(\theta)\rvert_{\theta=\pi} = \sum_{\epsilon} \frac{1}{L_x L_y} \frac{e^{\beta\epsilon}}{(e^{\beta\epsilon}-1)^2}
\end{aligned}
\label{F2pi}
\end{equation}
In the cylinder geometry, the hybridization between the two topological edge states is exponentially suppressed with the cylinder's circumference, causing the energy gap to decay as $\epsilon \sim e^{-L_y}$. With a fixed $L_y$, we denote the lowest-energy edge mode as $\epsilon(L_y,k^{0}_x)$. From \Eq{F2pi}, the contribution to $F^{(2)}(\theta)\big|_{\theta=\pi}$ is dominant by the edge modes with nearly zero energy. When $L_y$ is sufficiently large, the edge-mode energies collectively approach zero as $\epsilon(L_y,k^{0}_x) \sim \frac{v_F}{L_x}$, due to the finite size along the $x$ direction, where $v_F$ is the edge state velocity. The detailed calculation (SM) shows that the contribution from the edge modes to $F^{(2)}(\pi)$ follows the asymptotic behavior:
\begin{equation}
\begin{aligned}
F^{(2)}(\pi) \sim \frac{1}{(v_F \beta)^2}\frac{L_x}{ L_y}.
\end{aligned}
\label{ls}
\end{equation}
By subtracting the value at a reference length $L_x^0$, we define
the TSI, which isolates the relevant system-size dependence:
\begin{equation}
\mathcal{F}(\Delta L_x) = \beta^2\big(F_{L_x}^{(2)}(\pi) - F^{(2)}_{L_x^0}(\pi)\big),
\label{calf}
\end{equation}
where $\Delta L_x = L_x - L_x^0$. After subtracting the constant bulk contribution and the initial edge offset,  this quantity should be directly proportional to $\Delta L_x$, yielding a line through the origin with slope $1/(v_F^2 L_y)$. Our numerical results in \Fig{Fig1}(c) and (d) perfectly confirm this behavior. Consequently, the scaling behavior of the second derivative of $F(\theta)$ at $\theta=\pi$ establishes a feasible and reliable quantitative approach to characterize mixed-state topological phases.

{\em Kane-Mele-Hubbard model}--- We have introduced a framework for identifying topology in mixed states through the singular property in disorder operator. Next, through studying a prototypical interacting SPT model, namely the KMH model,  we demonstrate that our proposed linear scaling relation in \Eq{ls} persists even in the presence of strong interactions. This establishes the relation as a robust and effective tool for investigating mixed-state topology phases in interacting systems. To investigate the finite-temperature properties of interacting fermionic systems, we employ unbiased QMC simulations~\cite{AssaadReview,Li2019review}. Our analysis centers on the second-order derivative of the disorder operator, $F^{(2)}(\pi)$, which is rigorously expressed as:
$F^{(2)}(\pi)=\frac{1}{L^d} \frac{\langle (\text{i} \hat{Q})^2 e^{\text{i}\pi \hat{Q} }\rangle}{\langle e^{\text{i}\pi \hat{Q}} \rangle}$. 
Directly computing the ensemble average of the complex exponential operator $e^{i\pi Q}$ is numerically unstable and challenging within standard QMC frameworks. To overcome this, we have developed a novel and stable reweighting algorithm that enables an accurate evaluation of $F^{(2)}(\pi)$. The detailed description of our newly developed method is provided in the Supplemental Material.

We investigate the KMH model on a honeycomb lattice with the Hamiltonian given by
\bea
\hat{H} &=&\hat{H}_{\text{KM}}+\hat{H}_U \\
    \hat{H}_{\text{KM}} &=& -t\sum_{\langle i,j\rangle} \hat{c}^{\dagger}_{i\sigma} \hat{c}_{j\sigma} + \text{i} \lambda \sum_{\langle\langle i,j \rangle \rangle} \nu_{ij}  \hat{c}^{\dagger}_{i\alpha} \sigma^{z}_{\alpha\beta} \hat{c}_{j\beta}, \label{eq:KMH_KM} \\
    \hat{H}_{U} &=& U \sum_i (\hat{n}_{i\uparrow}-\frac{1}{2}) (\hat{n}_{i\downarrow} - \frac{1}{2}). \label{eq:KMH_U}
\eea
Here, $\hat{H}_{\text{KM}}$ is the standard Kane-Mele model with intrinsic spin-orbit coupling of strength $\lambda$~\cite{Kane2005PRLQSH}, and $\hat{H}_{U}$ introduces the Hubbard interaction of strength $U$. At $U=0$, the ground state is a TR-invariant topological insulator characterized by a $\mathbb{Z}_2$ topological invariant~\cite{Kane2010RMP,Zhang2011RMP}. At this point, the system hosts a pair of counter-propagating helical edge modes at the boundary, which cross at the TR-invariant point $k_x=\pi$, leading to a quantized spin Hall conductance—\text{i.e.}, the quantum spin Hall (QSH) effect. As $U$ increases, a quantum phase transition drives the system into an anti-ferromagnetic (AFM) Mott insulator, destroying the topological features~\cite{Wu2011PRBQSH,Vaezi2012PRB,Li2017PRBTI,Fiete2013PRB,Assaad2013Review}. Since the KMH model is sign-problem-free~\cite{Wu2005PRBsign,Li2015PRBsign,Li2016PRLsign,Xiang2016prl}, we can employ QMC to compute $F^{(2)}(\pi)$ and characterize its mixed-state topology across this transition. We fix $\lambda=0.1$, for which the ground-state transition is known to occur at $U_c \approx 5.0$~\cite{Assaad2012PRBQSH}. 

We first focus on the correlated topological phase in the regime ($U < U_c$), selecting $U=2.5$ as a representative interaction strength. Our simulations are performed on cylindrical geometries of size $L_x \times L_y$, with PBC in the $\hat{x}$-direction and OBC in the $\hat{y}$-direction to host edge states. At low temperatures ($T=1/5$), well below the bulk gap, we observe a distinct linear scaling of the TSI $\mathcal{F}$ with the change in cylinder circumference relative to the reference system, $\Delta L_x = L_x - L_x^0$, for various cylinder lengths $L_y$, as depicted in \Fig{Fig2}(a). Furthermore, the slope of this linear relation is found to be inversely proportional to $L_y$. This $\mathcal{F} \sim \frac{L_x}{L_y}$ behavior is the expected signature of helical edge states in a QSH insulator. Remarkably, this linear scaling persists up to relatively high temperatures, as shown in \Fig{Fig2}(b). The successful capture of these key topological signatures validates our numerical approach, establishing it as a reliable method for probing mixed-state topological properties in strongly correlated fermionic systems. 

We now employ our method to map out the interaction-driven topological phase transition. For interaction strengths below the critical point, such as $U=3.6$ and $U=4.5$, the TSI $\mathcal{F}$ exhibits a pronounced linear scaling with the relative system size $\Delta L_x$ [\Fig{Fig3}(a) and (b)]. This scaling is the definitive signature of the gapless helical edge states that characterize the QSH phase. In contrast, upon increasing the interaction to $U=6.0$, a qualitative change is observed. As shown in \Fig{Fig3}(c), the linear scaling vanishes, and $\mathcal{F}(\Delta L_x)$ becomes independent of $\Delta L_x$ and nearly vanishes, signaling a transition into a topologically trivial insulating phase. These contrasting behaviors clearly demarcate a phase transition from a QSH insulator to a trivial insulator. Our results are in agreement with previous ground-state sign-free QMC simulations that locate the quantum critical point at $U_c \approx 5.0$~\cite{Assaad2012PRBQSH}.

{\em Haldane-Hubbard model}---The Haldane model is a canonical example of a Chern insulator, demonstrating that a quantized Hall conductance arise in a lattice model without a net magnetic field. Unlike the Kane-Mele model, which preserves TR symmetry, the Haldane model explicitly breaks it, leading to the formation of robust chiral edge states. We investigate the interplay of topology and strong correlations by studying the spinful HH model, with a focus on its topological properties at finite temperature. The Hamiltonian is given by:
\begin{align}
    \hat{H} &= \hat{H}_{\text{Hal}} + \hat{H}_U \\
    \hat{H}_{\text{Hal}} &= -t \sum_{\langle i,j \rangle, \sigma} \hat{c}^{\dagger}_{i\sigma} \hat{c}_{j\sigma} + i\lambda \sum_{\langle\langle i,j \rangle\rangle, \sigma} \nu_{ij} \hat{c}^{\dagger}_{i\sigma} \hat{c}_{j\sigma} \\
    \hat{H}_{U} &= U \sum_i (\hat{n}_{i\uparrow}-\frac{1}{2}) (\hat{n}_{i\downarrow} - \frac{1}{2})
\end{align}
Here, $\hat{H}_{\text{Hal}}$ describes two independent copies of the Haldane model in \Eq{Haldane}, one for each spin species $\sigma \in \{\uparrow, \downarrow\}$. The Hubbard interaction is described by $\hat{H}_U$. The explicit TR symmetry breaking from the complex hopping term introduces a sign problem for any known QMC algorithm~\cite{Chang2024PRB,Yu2024arXiv,Mondaini2022science}. Consequently, despite its fundamental importance, unbiased numerical results for this model remain scarce~\cite{Troyer2016PRBHaldane,Zhai2015PRB,Troyer2016PRL,Hu2024PRB,Prokof2019PRB,Mondaini2021PRB}. However, the severity of the sign problem is known to diminish at higher temperatures~\cite{Troyer2005sign}. This thermal mitigation makes it feasible to apply our framework, based on DP, to obtain reliable results and systematically investigate the properties of the interacting mixed-state topological phase and its transition.

\begin{figure}[t]
    \centering
    \includegraphics[width=\columnwidth]{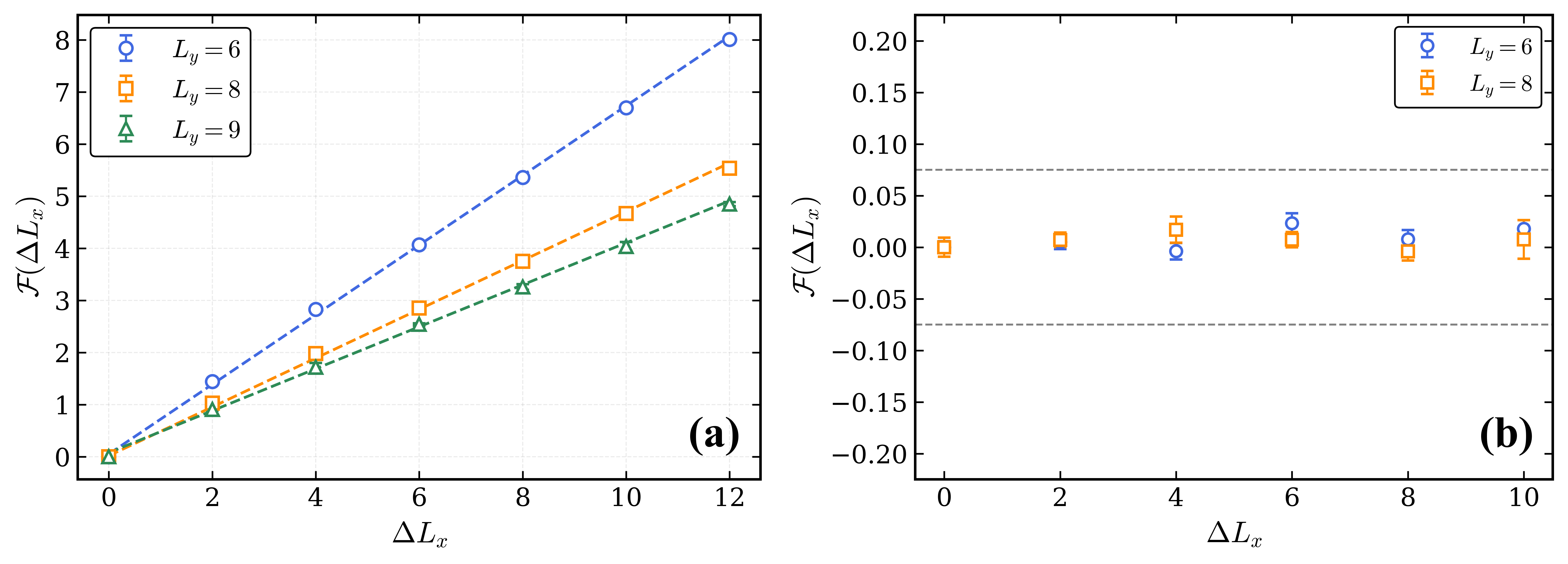}
    \caption{ The results of TSI in the HH model, with the inverse temperature fixed at $\beta = 3.6$ where the sign problem is still moderate. (a) $\mathcal{F}$ exhibits a clear linear relation with $\Delta L_x$, with the slope inversely proportional to $L_y$, indicating a topologically nontrivial phase.
(b) The linear relation disappears, and $\mathcal{F}$ nearly vanishes, suggesting a transition into a topologically trivial insulating phase. 
    }
    \label{Fig4}
\end{figure}

To investigate the stability of the topological phase against electron-electron interactions, we compute the TSI. In the weak-to-intermediate interaction regime (e.g. $U=2.5$ ), $\mathcal{F}$ exhibits a clear linear scaling with the relative system width, $\Delta L_x$, as shown in \Fig{Fig4}(a). This scaling is a distinct signature of the chiral gapless edge states, confirming that the topological phase remains robust. In contrast, at a strong interaction strength of $U=5$, a qualitative change occurs: the linear scaling of $\mathcal{F}$ vanishes, and its value extrapolates to zero in the thermodynamic limit (\Fig{Fig4}(b)). This behavior signals an interaction-driven quantum phase transition into a topologically trivial insulating phase. Our identification of a critical interaction strength near  $U_c \approx 5$ aligns well with previous studies establishing a first-order transition in this regime~\cite{Troyer2016PRBHaldane}.

{\em Concluding remarks}---In this work, we introduce a robust numerical framework for characterizing mixed-state topological phases in interacting fermionic systems. Our approach centers on the efficient QMC evaluation of $F^{(2)}(\pi)$, the second derivative of the DP generating function at a twist angle of $\pi$. We establish that in a topological phase, $F^{(2)}(\pi)$ exhibits a distinct linear scaling with system size $L_x$, serving as a direct probe of the gapless edge states. As a key demonstration, we apply this method to investigate the topological phase transition in the KMH model at finite temperature. Our simulations successfully map the interaction-driven transition from a QSH insulator to a trivial insulator, yielding results consistent with established ground-state studies. This work therefore establishes a robust and computationally efficient tool for exploring the rich landscape of topological phases in strongly correlated mixed states.

As a key application of our framework, we investigate the HH model, a canonical system whose study is notoriously hindered by a severe sign problem in QMC simulations. By performing simulations at finite temperatures where this sign problem is mitigated, we successfully characterize the model's mixed-state topological properties. Through the distinct scaling behavior of our indicator, $\mathcal{F}$, we identify the QAH phase, distinguished by its chiral edge states, and map out the interaction-driven phase transition to a topologically trivial insulator. Hence, our work establishes a viable new pathway to bypass the sign problem and quantitatively probe the topology of strongly correlated systems that were previously inaccessible.

{\em Acknowledgments.}----We appreciate helpful discussions with Ze-Min Huang. S.-H. S. and Z.-X. L. are supported by the National Natural Science Foundation of China under Grant Nos. 12347107 and 12474146, and Beijing Natural Science Foundation under Grant No. JR25007. X.-Q. S. acknowledges support from the Alexander von Humboldt foundation.

{\em Note added.}----Upon completing this work, we became aware of a related work by Ma, Xu and Jiang that also studies the the non-analytic behavior of the disorder parameter in 2D correlated topological insulators.\cite{Jiang2025arXiv}.

\bibliography{ref}
\bibliographystyle{apsrev4-1}


\clearpage
\onecolumngrid

\begin{center}
\textbf{\large Supplemental Material for\\[4pt]
``Diagnosis of mixed-state topological phases in strongly correlated systems via disorder parameters''}
\end{center}

\setcounter{equation}{0}
\renewcommand{\theequation}{S\arabic{equation}}

\setcounter{figure}{0}
\renewcommand{\thefigure}{S\arabic{figure}}

\setcounter{table}{0}
\renewcommand{\thetable}{S\arabic{table}}

This Supplementary Material provides additional details and supporting evidence for the conclusions presented in the main text. The document is organized into four sections. Section~I introduces the disorder parameter (DP) and presents the formulas for the derivatives of its logarithmic generating function used throughout this work. Section~II details our Quantum Monte Carlo algorithm, specifically explaining how these derivatives are computed in a numerically stable and efficient manner for interacting systems. In Sec.~III, we account for finite-size effects to analytically derive the singular behavior of the DP and its relevant derivatives at a twist angle $\theta = \pi$ in both one and two dimensions, driven by topological edge modes. Furthermore, we confirm the validity of Eq.~\eqref{ls} from the main text and investigate an additional model to demonstrate its universality and robustness. Finally, Sec.~IV outlines critical considerations for applying Eqs.~\eqref{ls} and \eqref{calf} to interacting systems and provides practical strategies to address them.

\section{Section~I.~Disorder operator and Derivatives of Its Logarithmic Generating Function}
In this section, we introduce the DP and its generating function. Under certain symmetry constraints, we present simplified analytic expressions for the derivatives of the generating function that will be used in the following.
DP provides a theoretical framework for describing complex charge fluctuations, and it is defined as follows:
\begin{equation}
Z(\theta) = \left\langle e^{i \theta \hat{Q}} \right\rangle ,
\end{equation}
where
\( \hat{Q} = \sum_{i,\sigma} \left( \hat{a}^{\dagger}_{i\sigma} \hat{a}_{i\sigma} - \frac{1}{2} \right)  \)
is the conserved charge corresponding to the $\mathrm{U}(1)$ symmetry.
The function $Z(\theta)$ is $2\pi$-periodic, so the range of $\theta$ can be restricted to the interval $[0, 2\pi)$. 
In a Hermitian system, it satisfies the relation $Z(\theta) = Z^{*}(-\theta)$.  Here, we are mainly interested in the associated generating function $F(\theta)$, defined as
\begin{equation}
    F(\theta) = \frac{1}{L^d} \Re\!\left[ \ln Z(\theta) \right]  = \frac{1}{L^d} \ln \left| Z(\theta) \right| ,
\end{equation}
where $L^d$ denotes the system size. It has been widely used to characterize phase transitions, symmetry breaking, and topological properties in quantum many-body systems.
In this work, we will show that the singular behavior of 
$F(\theta)$ at $\theta=\pi$ serves as an indicator for detecting the topology of mixed states. To capture this singular behavior, it is necessary to evaluate the derivatives of $F(\theta)$. We first present the formulas required for these calculations.
\begin{equation}
F^{(1)}(\theta) \;=\; \frac{1}{L^d}\,\Re\!\left[ \frac{1}{Z(\theta)} 
    \frac{\partial Z(\theta)}{\partial \theta} \right]
\;=\; \frac{1}{L^d}\,\Re\!\left[ \frac{\langle i \hat{Q} e^{i\theta \hat{Q}}\rangle}{\langle e^{i\theta \hat{Q}}\rangle} \right],
\label{eq:F1}
\end{equation}

\begin{equation}
F^{(2)}(\theta) \;=\; \frac{1}{L^d}\,\Re\!\left[
\frac{\langle (i\hat{Q})^2 e^{i\theta \hat{Q}}\rangle\langle e^{i\theta \hat{Q}}\rangle
- \langle i\hat{Q} e^{i\theta \hat{Q}}\rangle^2}
{\langle e^{i\theta \hat{Q}}\rangle^2}
\right].
\label{eq:F2}
\end{equation}
In a Hermitian system that is invariant under the particle-hole transformation, for example 
$\hat{c}_i \rightarrow (-1)^i \hat{c}_i^{\dagger}$ ($\hat{Q} \rightarrow -\hat{Q}$), the quantity 
$\langle (i\hat{Q})^{n} e^{i\theta \hat{Q}} \rangle$ (with $n=0$ corresponding to 
$Z(\theta)$) is necessarily real, since the transformation maps it to its 
Hermitian conjugate partner, $\langle (-i\hat{Q})^{n} e^{-i\theta \hat{Q}} \rangle$. 
In this case,  
Eqs.~\eqref{eq:F1} and \eqref{eq:F2} simplify to
\begin{equation}
F^{(1)}(\theta) = \frac{1}{L^d}\,
\frac{\langle i \hat{Q}\, e^{i\theta \hat{Q}}\rangle}{\langle e^{i\theta \hat{Q}}\rangle},
\label{eq:F1_simplified}
\end{equation}
and
\begin{equation}
F^{(2)}(\theta) = \frac{1}{L^d}\,
\frac{\langle (i\hat{Q})^2 e^{i\theta \hat{Q}}\rangle\langle e^{i\theta \hat{Q}}\rangle
- \langle i\hat{Q} e^{i\theta \hat{Q}}\rangle^2}
{\langle e^{i\theta \hat{Q}}\rangle^2}.
\label{eq:F2_simplified}
\end{equation}
In particular, we will mainly focus on  $\theta = \pi$, where a singularity may occur. Since $Z(\theta) = Z^{*}(-\theta)$ and $Z(\theta)$ is real, 
$Z(\theta)$ is a real even function with period $2\pi$, so it is symmetric 
about $\theta=\pi$, and all of its odd derivatives must vanish at $\theta=\pi$.
\begin{equation}
    \left.\frac{\partial^{\,2k+1} Z(\theta)}{\partial \theta^{\,2k+1}}\right|_{\theta=\pi} = \left.\langle (i\hat{Q})^{2k+1} e^{i\theta \hat{Q}}\rangle\right|_{\theta=\pi} =  0 ,
\end{equation}
Therefore we can conclude
\[
F^{(1)}(\pi)=0,
\qquad
F^{(2)}(\pi)=\left.\frac{1}{L^d}\,
\frac{\langle (i\hat{Q})^2 e^{i\theta \hat{Q}}\rangle}{\langle e^{i\theta \hat{Q}}\rangle}\right|_{\theta=\pi}.
\]
However, in more general settings,  one may be concerned not with the total charge $\hat{Q}$ of the full system, but with the charge $\hat{Q}_b$ associated with a subsystem, such as the boundary region. In this case, as long as the particle--hole transformation still maps $\hat{Q}_b \to -\hat{Q}_b$, all of the above formulas continue to hold.

\section{Section~II.~Determinant Quantum Monte Carlo Algorithm}
In this section, we detail the determinant quantum Monte-Carlo (DQMC) algorithm developed to compute the derivatives of $F(\theta)$ reliably and robustly. As established in Sec.~I, evaluating these derivatives requires computing the ratio
\[
\frac{\langle (i \hat{Q})^n e^{i\theta \hat{Q}} \rangle}{\langle e^{i\theta \hat{Q}} \rangle}.
\] 
In a sign-problem-free system, the most straightforward approach is to compute 
$\langle (i \hat{Q})^n e^{i\theta \hat{Q}} \rangle$ and $\langle e^{i\theta\hat{Q}} \rangle$ separately 
and then take their ratio. However, since $\langle e^{i\theta \hat{Q}} \rangle$ decays exponentially with system size, accurate evaluation is difficult. Furthermore, the vanishingly small denominator amplifies statistical errors, necessitating excessive computational effort to achieve reliable results.

As is well known, the computation of the free energy requires evaluating the partition function, which grows exponentially with system size and thus poses numerical difficulties. Similarly, here computing $F(\theta) = \frac{1}{L^d}\ln|\langle e^{i\theta \hat{Q}}\rangle|$ requires evaluating $\langle e^{i\theta \hat{Q}}\rangle$, which faces analogous numerical instabilities in DQMC. Nevertheless, derivatives of the free energy—which correspond to various physical observables—can be accurately obtained using quantum Monte Carlo methods. The same idea applies here: although 
$\langle e^{i\theta \hat{Q}}\rangle$ cannot be computed directly, the derivatives of $F(\theta)$, or equivalently the ratio 
$\frac{\langle (i \hat{Q})^n e^{i\theta \hat{Q}}\rangle} {\langle e^{i\theta \hat{Q}} \rangle}$ can be 
reliably evaluated. Here, we introduce an efficient and numerically stable algorithm to precisely compute the ratio.

Firstly, we note that
\begin{equation}
\frac{\langle (i \hat{Q})^n e^{i\theta \hat{Q}} \rangle}{\langle e^{i\theta \hat{Q}} \rangle}
= \frac{\mathrm{Tr}\big(e^{-\beta \hat{H}} (i \hat{Q})^n e^{i\theta \hat{Q}}\big)}
       {\mathrm{Tr}\big(e^{-\beta \hat{H}} e^{i\theta \hat{Q}}\big)}.
\end{equation}
Since $\hat{Q}$ is the generator of the U(1) symmetry, it commutes with the Hamiltonian, $[\hat{H}, \hat{Q}] = 0$, and we can introduce a non-Hermitian effective Hamiltonian
\begin{equation}
\hat{H}_{\rm eff} = \hat{H} - \frac{i\theta}{\beta} \hat{Q},
\end{equation}
so that the above ratio can be written as an ensemble average over the non-Hermitian Hamiltonian $\hat{H}_{\rm eff}$:
\begin{equation}
\frac{\langle (i \hat{Q})^n e^{i\theta \hat{Q}} \rangle}{\langle e^{i\theta \hat{Q}} \rangle}
= \frac{\mathrm{Tr}\big(e^{-\beta \hat{H}_{\rm eff}} (i \hat{Q})^n \big)}
       {\mathrm{Tr}\big(e^{-\beta \hat{H}_{\rm eff}}\big)}.
\end{equation}
Now, we can fully adopt the traditional finite-temperature DQMC~\cite{AssaadReview} algorithm to simulate a system with the equivalent Hamiltonian 
$\hat{H}_{\rm eff} = \hat{H}_0 + \hat{H}_{\mathrm{int}}$, where 
$\hat{H}_0 = \hat{H}_{\mathrm{f}} - \frac{i\theta}{\beta} \hat{Q}$ with $\hat{H}_{\mathrm{f}}$ being the quadratic one-body term and $\hat{H}_{\mathrm{int}}$ being the quartic two-body interaction term in the original physical Hamiltonian $\hat{H}$.

To elucidate the technical details of the algorithm, we next take the
Kane--Mele--Hubbard model studied in the main text as a concrete example.
In this case, $\hat{H}_0 = \hat{H}_{\mathrm{KM}}-\frac{i\theta\hat{Q}}{\beta}$, and the interaction part is simply the Hubbard term $\hat{H}_{\mathrm{int}} = \hat{H}_U$. 
As usual, we first employ a symmetric Trotter decomposition to separate $\hat{H}_0$ and $\hat{H}_U$:
\begin{equation}
\mathcal{Z}_{\rm eff} = \mathrm{Tr}\, e^{-\beta \hat{H}_{\rm eff}}
= \mathrm{Tr} \left[ \left( e^{-\frac{\Delta\tau}{2} \hat{H}_0} e^{-\Delta\tau \hat{H}_U} e^{-\frac{\Delta\tau}{2} \hat{H}_0} \right)^m \right] + \mathcal{O}(\Delta\tau^3),
\end{equation}
where $m \Delta \tau = \beta$, and the inverse temperature $\beta$ is divided into $m$ slices. We then apply the Hubbard-Stratonovich (HS) transformation 
to decouple the interaction term , introducing 
an auxiliary field $s_l(i)$ at each site $i$ and time slice $l$:
\begin{equation}
e^{-\Delta\tau U (\hat{n}_{i\uparrow}-\frac{1}{2})(\hat{n}_{i\downarrow}-\frac{1}{2})} 
= C \sum_{s_l(i)=\pm 1} 
e^{i \alpha s_l(i) (\hat{n}_{i\uparrow}+\hat{n}_{i\downarrow}-1)},
\end{equation}
where $\cos(\alpha) = e^{-\Delta\tau U/2}$ and $C$ is a constant. After performing the HS decoupling, the fermionic degrees of freedom become quadratic and 
can be traced out analytically. As a result, the partition function can be expressed as a sum over the weights 
of all possible configurations of the auxiliary field:
\begin{equation}
\mathcal{Z}_{\rm eff} = C^m \sum_{\{s_l(i) = \pm 1\}} 
\mathrm{Tr} \left[ \prod_{l=1}^{m} 
 e^{-\frac{\Delta\tau}{2} \hat{H}_0} \, e^{i \alpha \sum_i s_l(i) (\hat{n}_{i\uparrow}+\hat{n}_{i\downarrow}-1)} 
\, e^{-\frac{\Delta\tau}{2} \hat{H}_0} \right] = \sum_{\mathbf{s}} W_\mathbf{s}\ ,
\end{equation}
where $\mathbf{s} = \{ s_l(i) = \pm 1 \}$ denotes the set of auxiliary fields. For the Kane–Mele–Hubbard model considered here, we can prove that the configuration weights $W_{\mathbf{s}}$
are strictly non-negative.
Under the transformation
$\hat{c}_{i\downarrow} \rightarrow (-1)^{i} \hat{c}^{\dagger}_{i\downarrow}$, the fermionic trace 
factorizes into two complex-conjugate parts. Explicitly,
\begin{equation}
\mathrm{Tr}\!\left[
\prod_{l=1}^{m}
e^{-\frac{\Delta\tau}{2} \hat{H}_{0}}\,
e^{i\alpha \sum_{i} s_{l}(i)\,(\hat{n}_{i\uparrow}+\hat{n}_{i\downarrow}-1)}\,
e^{-\frac{\Delta\tau}{2} \hat{H}_{0}}
\right]
\;\longrightarrow\;
\mathrm{Tr}\!\left[
\prod_{l=1}^{m}
e^{-\frac{\Delta\tau}{2} (\hat{H}^{\uparrow}_0+\hat{H}^{\downarrow}_0)}\,
e^{i\alpha \sum_{i} s_{l}(i)\,(\hat{n}_{i\uparrow}-\hat{n}_{i\downarrow})}\,
e^{-\frac{\Delta\tau}{2} (\hat{H}_0^{\uparrow}+\hat{H}_0^{\downarrow})}
\right].
\end{equation}

where
\begin{align}
\hat{H}_0^{\uparrow} &= \hat{H}_{\mathrm{KM}}^{\uparrow} - \frac{i\theta}{\beta}\sum_{i} \hat{n}_{i\uparrow},\\
\hat{H}_0^{\downarrow} &= \hat{H}_{\mathrm{KM}}^{\downarrow} + \frac{i\theta}{\beta}\sum_{i} \hat{n}_{i\downarrow},
\end{align}
and, $\hat{H}_{\mathrm{KM}}$ is invariant under the transfromation, the
spin-up $\hat{H}_{\mathrm{KM}}^{\uparrow}$ and spin-down $\hat{H}_{\mathrm{KM}}^{\downarrow}$ sectors are related by the TR symmetry,  hence also $H^{\uparrow}_0 = {H^{\downarrow}_0}^{*}$ in real space.
Consequently, the
configuration weight $W_{\mathbf{s}}$ factorizes into a product of two complex-conjugate parts, and is thus strictly non–negative. Finally, an expression of the form $\langle \hat{O} \rangle_{H_{\mathrm{eff}}} = \frac{\mathrm{Tr}\!\left( \hat{O}\, e^{-\beta \hat{H}_{\rm eff}} \right)}
 {\mathrm{Tr}\!\left( e^{-\beta \hat{H}_{\rm eff}} \right)}$, can be rewritten as a weighted average over auxiliary–field configurations:
 \begin{equation}
\langle O \rangle
= \frac{\sum_{\mathbf{s}}\, O_{\mathbf{s}} W_{\mathbf{s}} }
       {\sum_{\mathbf{s}} W_{\mathbf{s}} } \, .
 \end{equation}
   Thus, we can construct a Markov chain that samples the auxiliary–field configurations
according to their weights $W_{\mathbf{s}}$, using the standard Metropolis update scheme. 
However, if we are concerned only with the charge $\hat{Q}_b$ of a subregion, 
the fact that $[\hat{Q}_b, \hat{H}]\neq 0$ prevents us from defining such a non-Hermitian 
effective Hamiltonian $H_{\mathrm{eff}}$ as in the full-system case. Nevertheless, the effective partition function
\[
\mathcal{Z}_{\mathrm{eff}} = \mathrm{tr}\!\left( e^{-\beta \hat{H}} e^{i\theta \hat{Q}_b} \right)
\]
remains well defined, and it can still be decomposed into a sum of weights $W_{\mathbf{s}}$ over all possible configurations after HS transformation.
\begin{equation}
\mathcal{Z}_{\rm eff}
= C^m \sum_{\{s_l(i)=\pm1\}}
\mathrm{Tr}\!\left[
\left(\prod_{l=1}^{m}
e^{-\frac{\Delta\tau}{2}\hat{H}_f}\,
e^{i\alpha\sum_i s_l(i)\,(\hat{n}_{i\uparrow}+\hat{n}_{i\downarrow}-1)}\,
e^{-\frac{\Delta\tau}{2}\hat{H}_f}
\right)
e^{i\theta \hat{Q}_b}
\right]
= \sum_{\mathbf{s}} W_{\mathbf{s}} \,,
\end{equation}
where $\hat{Q}_b = \sum_{i \in b,\sigma}(\hat{c}^{\dagger}_{i\sigma}\hat{c}_{i\sigma}-\frac{1}{2})$. For the Kane--Mele Hubbard model, the same complex--conjugate factorization method can be used to prove that the weights $W_{\mathbf{s}}$ are still non--negative.

Unfortunately, the weights $W_{\mathbf{s}}$ are not always positive, as in, for example, the Haldane–Hubbard model, giving rise to the notorious sign problem. To circumvent this, one samples the auxiliary–field configurations according to $|W_{\mathbf{s}}|$, and then recovers an unbiased estimate of $\langle  \hat{O} \rangle$ by introducing the average sign of $W_{\mathbf{s}}$.
\begin{equation}
\langle \hat{O} \rangle
= \frac{\sum_{\mathbf{s}} O_{\mathbf{s}} \, \mathrm{sign}(W_{\mathbf{s}}) \, |W_{\mathbf{s}}|}
       {\sum_{\mathbf{s}} \mathrm{sign}(W_{\mathbf{s}}) \, |W_{\mathbf{s}}|}
       =\frac{ \langle O_{\mathbf{s}} \, \mathrm{sign}(W_{\mathbf{s}}) \rangle_{\left| W_{\mathbf{s}} \right|}       }{{\langle \mathrm{sign}(W_{\mathbf{s}})\rangle}_{\left| W_{\mathbf{s}} \right|}}
\end{equation}
In general, the average fermion sign in the denominator decays exponentially with system size and inverse temperature. Consequently, obtaining reliable results requires an exponentially increasing computational effort. Nevertheless, the topological indicator $\mathcal{F}$ proposed in this work remains robust even at relatively high temperatures, allowing us to investigate models plagued by severe sign problems that were previously inaccessible with conventional approaches.

\section{Section~III.~Disorder Parameters in Free-Fermion Systems}
To illustrate the fundamental behavior of the disorder parameter in free-fermion systems, we first analyze a single fermionic mode. Its density operator $\hat{\rho}$, in thermal equilibrium at inverse temperature $\beta$, is given by
\begin{equation}
\hat{\rho} = \frac{e^{-\beta\varepsilon\, (\hat{c}^{\dagger} \hat{c} - \frac{1}{2})}}
           {2\cosh(\frac{\beta\varepsilon}{2})},
\label{eq:rho_single_mode}
\end{equation}
where $\varepsilon$ denotes the single-particle energy.
  This simplified framework allows for the disorder parameter to be evaluated analytically. It is straightforward to show that
\begin{equation}
f_{\varepsilon}(\theta)
= \frac{1}{L^d}
  \ln \sqrt{1-\frac{\sin^2(\theta/2)}{\cosh^2(\beta\varepsilon/2)} }
  ,
\label{eq:F_single_mode}
\end{equation}
and its first derivative is
\begin{equation}
f^{(1)}_{\varepsilon}(\theta)
= - \frac{1}{2L^d}\,
    \frac{ 1 }
         { \tan(\theta/2) \sinh^2(\beta\varepsilon/2) + \frac{\cosh^2(\beta\varepsilon/2)}{\tan(\theta/2)}    }   .
\label{eq:F1_single_mode}
\end{equation}
From Eq.~\eqref{eq:F_single_mode}, one can already see that
a zero mode with $\varepsilon = 0$ can induce a singular behavior
of $f_{\varepsilon}(\theta)$ at $\theta = \pi$.
For a general free-fermion system, $F(\theta)$ decomposes into a sum of contributions from all single-particle eigenmodes:
\begin{equation}
F(\theta)=\sum_{\varepsilon} f_{\varepsilon}(\theta).
\end{equation}
In the following, we analyze the one- and two-dimensional cases individually. We demonstrate how the singular behavior of $F(\theta)$ at $\theta=\pi$, which originates from zero-energy edge modes, manifests in finite-size systems. Furthermore, we establish how this singularity serves as a robust and practical indicator for diagnosing mixed-state topological phases.

\subsection{Section~III-A.~Free Fermions in One Dimension}
In one-dimensional topological phases, the edge modes emerge as exponentially localized end states.
Due to the finite system size, their energies acquire an exponentially small splitting,  
\(\varepsilon_0 \sim e^{-L/\ell}\), where \(\ell\) is a characteristic decay length.  
We now perform a careful finite-size analysis to examine the asymptotic singular behavior of $F(\theta)$ in the vicinity of $\theta=\pi$,which is governed by the zero-energy edge modes.

\begin{figure*}[htbp]
    \centering
    \includegraphics[width=0.95\textwidth]{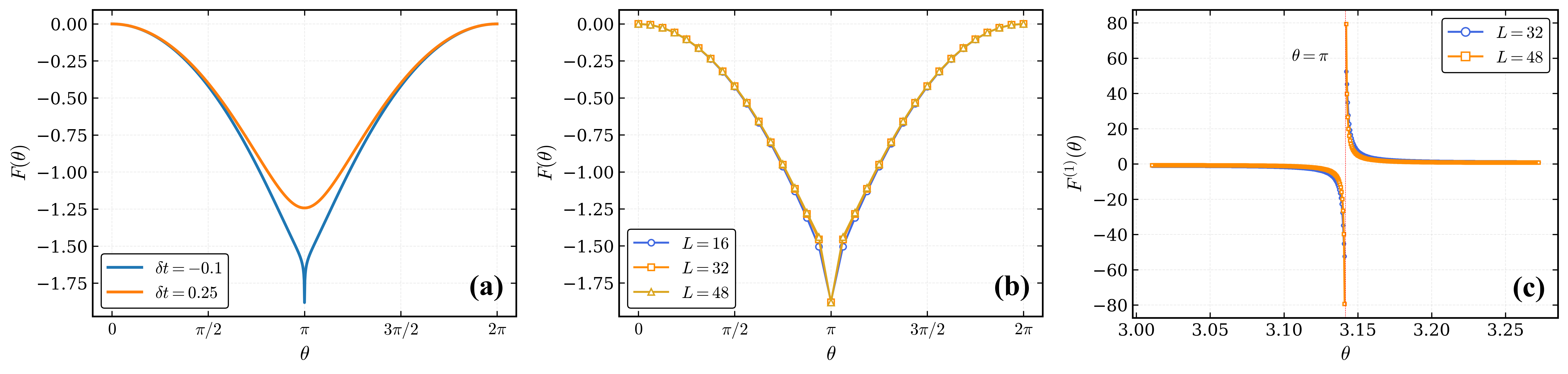}
    \caption{
(a) $F(\theta)$ in the trivial ($\delta t=0.25$) and topological ($\delta t=-0.1$) regimes, showing a cusp only in the latter.
(b) $F(\theta)$ in the topological regime ($\delta t=-0.1$) for various system sizes $L$.
(c) $F^{(1)}(\theta)$ in the topological regime ($\delta t=-0.1$) for different $L$, exhibiting a divergence at $\theta=\pi$ as $L$ increases.
The inverse temperature is set to $\beta = 1.0$ throughout.
    }
    \label{fig:S1}
\end{figure*}

At $\theta=\pi$, the contribution of the zero-energy edge modes to 
$F(\theta)$ is given by
\begin{equation}
f_{\varepsilon_0}(\pi)
    = \frac{1}{L} 
      \ln \sqrt{
      1 - \frac{1}{\cosh^2(\beta\varepsilon_0/2)}
      } = \frac{1}{L} \ln \left| \tanh\!\left( \frac{\beta\varepsilon_0}{2} \right) \right|.
\label{eq:f_edge_pi}
\end{equation}
The leading contribution $f_{\varepsilon_0}(\pi) \sim 1/\ell$ is always finite, and hence $F(\theta)$ remains 
continuous at $\theta=\pi$ with increasing $L$. We then turn our attention to $F^{(1)}(\theta)$. In the vicinity of $\theta = \pi$, the magnitude of $f^{(1)}_{\varepsilon_0}(\theta)$ attains its maximum
\begin{equation}
    \left| f^{(1)}_{\varepsilon_0}(\theta)\right| \leq \frac{1}{2L} \frac{1}{2\sqrt{\tan(\theta/2) \sinh^2(\beta\varepsilon_0/2)  \frac{\cosh^2(\beta\varepsilon_0/2)}{\tan(\theta/2)}}} =  \frac{1}{2L\sinh(\beta\varepsilon_0)} \sim \frac{e^{\frac{L}{\ell}}}{\beta L}
\end{equation}
at $\left| \tan\frac{\theta_0}{2} \right| = \left| \coth\frac{\beta\varepsilon_0}{2} \right|$, 
which is equivalent to $\left| \tan \theta_0 \right| = \left| \sinh \beta\varepsilon_0 \right|$ and $\left| \theta_0 - \pi \right| \approx \beta e^{-L/\ell}$. This indicates that, near $\theta = \pi$, $F^{(1)}(\theta)$ diverges exponentially with the system size, 
leading to a sharp cusp in $F(\theta)$ at $\theta = \pi$. However, in the topologically trivial phase, no zero-energy edge modes are present, 
and both $F(\theta)$ and $F^{(1)}(\theta)$ remain smooth functions.
 For concreteness, we illustrate the above discussion using the one-dimensional SSH model introduced in the main text. The corresponding results are shown in Fig.~\ref{fig:S1}, where panel~(a) compares the behavior of
$F(\theta)$ in the trivial and topological regimes, panel~(b) presents $F(\theta)$ in the
topological phase for different system sizes, and panel~(c) shows the divergence of
$F^{(1)}(\theta)$ at $\theta=\pi$ as $L$ increases.

\subsection{Section~III-B.~Free Fermions in two Dimension}
For the two-dimensional case, the situation is slightly different. We keep the $x$ direction periodic while opening the $y$ direction, which generates a boundary supporting edge states. Accounting for the anisotropy between the $x$ and $y$ directions, we have already replaced $L^d$ by $L_x L_y$. In this geometry the crystal momentum $k_x$ remains a good quantum number, and the single-particle Hamiltonian decomposes into a direct sum of one-dimensional blocks, $H = \bigoplus_{k_x} H(k_x)$. Now the edge modes are no longer isolated end states as in the one-dimensional case. Instead, they form a continuous dispersion with respect to $k_x$, giving rise to an edge band traversing the bulk gap. Generally, due to finite-size effects, the edge-mode energies typically vanish as $\Lambda_{\min} \sim \frac{v_F}{L_x}$, with $v_F$ denoting the Fermi velocity. Here we need to take into account all edge states in the vicinity of the Fermi energy $E_F = 0$. We adopt a linear dispersion approximation, where the energy spacing is given by $\frac{2\pi v_F}{L_x}$ , and introduce an energy cutoff $\Lambda_{\rm max}$. The contribution to $F^{(1)}(\theta)$ is then given by
\begin{equation}
    \left| \sum_{\Lambda_{\min}<\varepsilon<\Lambda_{\max}}  f^{(1)}_{\varepsilon}(\theta) \right | 
    \leq   \sum_{\Lambda_{\min}<\varepsilon<\Lambda_{\max}}  \frac{1}{2L_xL_y \sinh(\beta\varepsilon)}  
    \sim \frac{1}{L_y\beta v_F} \int_{\Lambda_{\min}}^{\Lambda_{\max}} \frac{d(\beta\varepsilon)} {\beta\varepsilon} 
    \sim \frac{1}{(v_F\beta)^2}\frac{\ln (L_x/ v_F\beta)}{L_y/ v_F\beta} \,,
\end{equation}
which vanishes as both $L_x$ and $L_y$ increase simultaneously.
 Therefore, in general, $F(\theta)$ remains smooth even in the topological phase, exhibiting no cusp. However, in certain special cases, the momentum point $k_x^0$ at which the zero-energy mode appears in the thermodynamic limit happens to coincide with one of the discrete $k_x$ points. Since its energy decays exponentially with the open-direction length $L_y$ as
$
\varepsilon(k_x^0, L_y) \sim e^{-L_y/\ell_y},
$
this restores the divergence of $F^{(1)}(\pi)$ and produces a cusp in $F(\theta)$ at $\theta=\pi$. 
This observation motivates us, in general case, to consider threading a magnetic flux through the cylinder that evolves continuously as $2\pi \tau$ from $\tau = 0$ to $1$, which effectively shifts $k_x$ to $k_x + 2\pi/L_x$. At some intermediate $\tau$, the momentum $k_x^0$ of the zero-energy mode necessarily coincides with one of these discrete $k_x$ points, which also produces a sharp cusp in the curve of $F|_{\theta=\pi}(\tau)$, as illustrated in Fig.~\ref{Fig1}(b) of the main text.

\begin{figure}[t] 
    \centering
    \includegraphics[width=\linewidth]{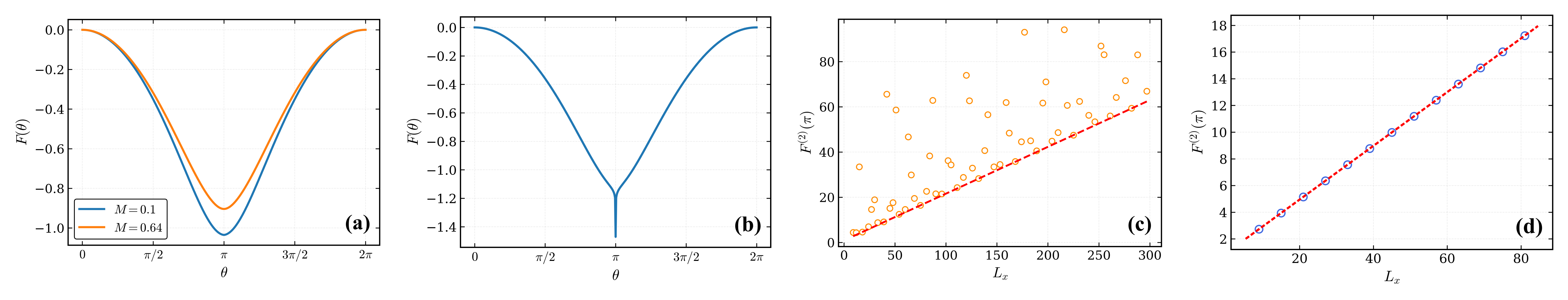}  
    \caption{
All calculations are carried out for the Haldane model with
\(t=1.0,\; \lambda=0.1\). The system is a topological phase for \(|M|<3\sqrt{3}\lambda\approx0.5196\). (a)
\(F(\theta)\) computed in the topological($M=0.10$) and trivial($M=0.64$) phases with $L_x=L_y=64$ — in both
cases there is no cusp and \(F(\theta)\) is smooth. (b) For \(M=0\), the zero-energy edge mode intersects the Fermi level exactly at \(k_x^{0}=\pi\); a
pronounced cusp appears with \(L_x=L_y=8\). (c) In the topological phase ($M=0.1$),
\(F^{(2)}(\pi)\) at fixed \(L_y=8\) shows an oscillatory linear
dependence on \(L_x\), and its lower envelope forms a straight line. (d) For \(M=0\) and fixed \(L_y=8\), odd \(L_x\)
yields an almost perfectly linear dependence of \(F^{(2)}(\pi)\) on
\(L_x\). The inverse temperature is set to $\beta = 1.0$ throughout.
    }
    \label{fig:S2}
\end{figure}

Next, we will show that even in the absence of a threaded magnetic flux, the scaling behavior of $F^{(2)}(\pi)$, originating from the edge modes, by itself provides an efficient tool to extract the mixed-state topology.
We now examine the contribution of the edge modes to $F^{(2)}(\theta)$ at $\theta=\pi$. 
\begin{equation}
    \sum_{\Lambda_{\min}<\varepsilon<\Lambda_{\max}} f^{(2)}_{\varepsilon}(\pi) = \sum_{\Lambda_{\min}<\varepsilon<\Lambda_{\max}} \frac{1}{4L_x L_y} \frac{1}{\sinh^2(\beta\varepsilon/2)}
\end{equation}
For the general case, we assume $L_y \sim \ell_y$ is sufficiently large so that $\varepsilon(k_x^0, L_y)$ can be regarded as zero, thereby ensuring that the asymptotic behavior $\Lambda_{\min} \sim \frac{v_F}{L_x}$ holds. Under these assumptions, we have
\begin{equation}
    \sum_{\Lambda_{\min}<\varepsilon<\Lambda_{\max}} f^{(2)}_{\varepsilon}(\pi) \sim  \frac{1}{L_y\beta v_F}\int_{\Lambda_{\min}}^{\Lambda_{\max}} \frac{d(\beta\epsilon)}{(\beta\epsilon)^2} \sim \frac{1}{(v_F\beta)^2} \frac{L_x}{L_y}.
\end{equation}
Now we obtain the final form 
\begin{equation}
    F^{(2)}_{\mathrm{edge}}(\pi) = \sum_{\Lambda_{\min}<\varepsilon<\Lambda_{\max}} f^{(2)}_{\varepsilon}(\pi) \sim \frac{1}{(v_F\beta)^2}\frac{L_x}{L_y}
    \label{S_ls}
\end{equation}
described in the main text. The bulk contribution $F^{(2)}_{\mathrm{bulk}}(\pi)$ is strongly suppressed by the gap and converges rapidly and can be treated as a constant. 
Therefore, for a fixed $L_y$, $F^{(2)}(\pi)$ exhibits an asymptotically linear dependence on $L_x$.
 Its asymptotic linear behavior with respect to $L_x$ can then be used to probe the topology of the mixed state.
In general, $F^{(2)}(\pi)$ exhibits an oscillatory linear scaling with $L_x$, or more precisely its lower envelope is a straight line, since $\Lambda_{\min}$ does not approach zero monotonically as $\frac{v_F}{L_x}$, but actually oscillates. However, in special cases, $\Lambda_{\min}$ can approach zero almost monotonically as $\frac{v_F}{L_x}$. For instance, when $k_x^0 = \pi$, we can choose $L_x$ to be odd, and $F^{(2)}(\pi)$ versus $L_x$ then forms an almost perfect straight line.  
It is thus natural to define, as in the main text,
$
\mathcal{F} = \beta^2 \big( F^{(2)}(\pi) - F_0^{(2)}(\pi) \big)
$. It then forms a straight line passing through the origin.
We illustrate the above discussion using the Haldane model introduced in the main text. The corresponding results are shown in Fig.~\ref{fig:S2}, 
where panel~(a) shows that $F(\theta)$ is generically smooth in both the topological and trivial phases, 
Panel~(b) demonstrates that for $M=0$, choosing an even $L_x$ ensures that $k_x^0 = \pi$ is included among the discrete $k_x$ values, resulting in the appearance of a cusp in $F(\theta)$ at $\theta=\pi$.
and panels~(c) and (d) together show that, at fixed $L_y$, $F^{(2)}(\pi)$ exhibits an asymptotically linear dependence on $L_x$.

\subsection{Section~III-C.~Qi--Wu--Zhang Model on Square Lattice}
To establish the universality of our framework and rule out model-specific artifacts, we extend our analysis to a second microscopic system: the Qi--Wu--Zhang (QWZ) model~\cite{qi2006prb}, which realizes a topological insulator on a square lattice. In momentum space, its Hamiltonian is given by
\begin{equation}
H(\vec{k}) = \lambda \sin(k_x)\sigma_x
           + \lambda \sin(k_y)\sigma_y
           + \bigl( 1 + \lambda\cos(k_x) + \lambda\cos(k_y) \bigr)\sigma_z .
\end{equation}

\begin{figure}[htbp] 
    \centering
    \includegraphics[width=\linewidth]{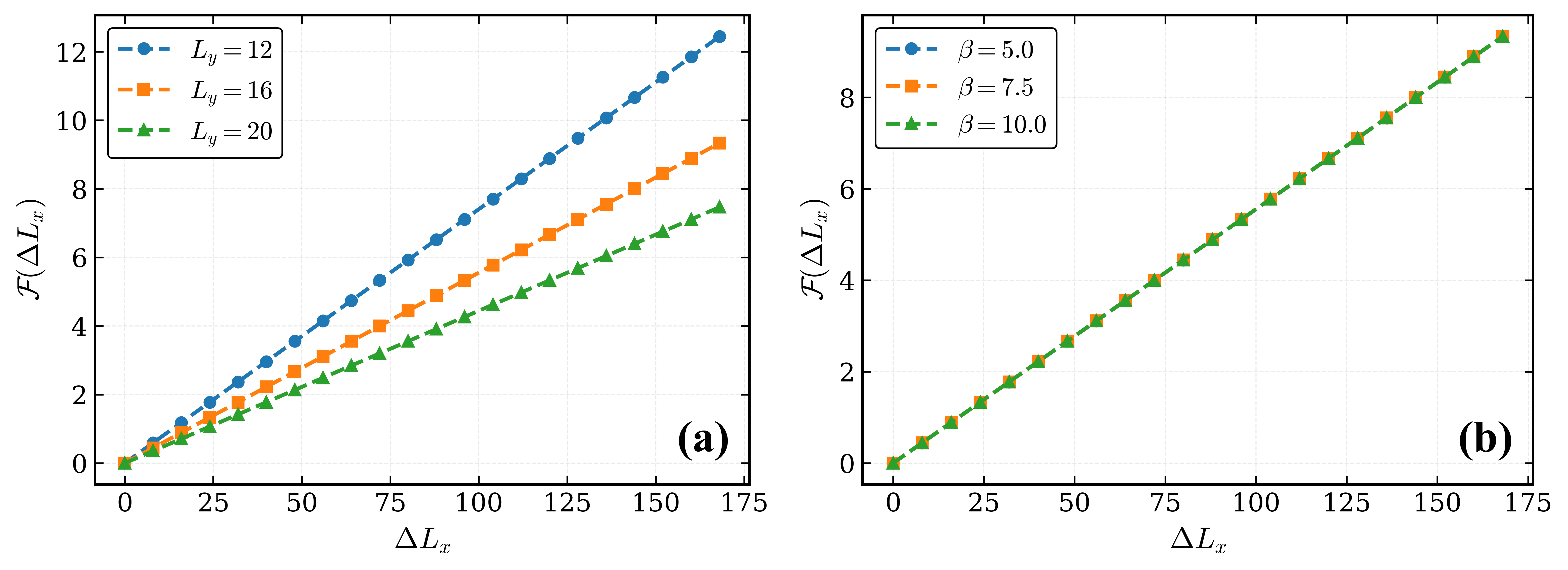}  
    \caption{Both panels are obtained at $\lambda = 0.75$, where the system lies in the topological phase. 
(a) For $\beta = 5.0$, the topological indicator $\mathcal{F}$ is evaluated at different $L_y$. 
The results show a clear linear dependence on $\Delta L_x$, with a slope that decreases as $L_y$ increases. 
(b) With $L_y$ fixed at $16$, $\mathcal{F}$ is computed for several temperatures. 
In this free–fermion limit, all curves still collapse onto the same straight line.
Throughout, the reference system size is taken as $L_x^{0}=19$.
    }
    \label{fig:S3}
\end{figure}

For $|\lambda| > \tfrac{1}{2}$, the system carries a nonzero Chern number.
By performing a Fourier transform, we obtain the Hamiltonian in real space as
\begin{equation}
\hat{H} = \hat{H}_0 + \lambda \hat{H}_1 + \lambda \hat{H}_2 ,
\end{equation}
where
\begin{align}
\hat{H}_0 &= \sum_{i,j} \bigl( \hat{a}^\dagger_{i,j} \hat{a}_{i,j}
      - \hat{b}^\dagger_{i,j} \hat{b}_{i,j} \bigr), \\
\hat{H}_1 &= \frac{1}{2}\sum_{i,j} \Bigl(
      \hat{a}^\dagger_{i+1,j} \hat{a}_{i,j} - \hat{b}^\dagger_{i+1,j} \hat{b}_{i,j}
      + \text{h.c.} \Bigr)
      + \frac{1}{2}\sum_{i,j} \Bigl(
      \hat{a}^\dagger_{i,j+1} \hat{a}_{i,j} - \hat{b}^\dagger_{i,j+1} \hat{b}_{i,j}
      + \text{h.c.} \Bigr), \\
\hat{H}_2 &= \frac{1}{2}\sum_{i,j} \Bigl(
      i \hat{a}^\dagger_{i+1,j} \hat{b}_{i,j} + i \hat{b}^\dagger_{i+1,j} \hat{a}_{i,j}
      + \text{h.c.} \Bigr)
      + \frac{1}{2}\sum_{i,j} \Bigl(
      \hat{a}^\dagger_{i,j+1} \hat{b}_{i,j}
      - \hat{b}^\dagger_{i,j+1} \hat{a}_{i,j}
      + \text{h.c.} \Bigr).
\end{align}
We again take the system to be periodic along $x$ and open along $y$. Within this setup, we show in Fig.~\ref{fig:S3} that the topological indicator $\mathcal{F}$ we defined continues to work robustly, i.e., $\mathcal{F}(\Delta L_x) \propto \frac{1}{L_yv_F^2} \Delta L_x$ in topological phase.

\section{Section~IV.~Additional quantum Monte-Carlo results in Two-Dimensional interacting systems}
In this section, we present additional QMC results for two-dimensional interacting systems, further corroborating the conclusions drawn in the main text. As noted in Sec.~III-B, the validity of the linear relation in Eq.~\eqref{S_ls} requires that the system dimension along the open boundary, $L_y$, be comparable to the characteristic length scale $\ell_y$. In interacting systems, however, $\ell_y$ is generally unknown \emph{a priori} and depends on both microscopic parameters and thermal fluctuations (via $\beta$). We illustrate this using the KMH model. We find that $\ell_y(U,\beta)$ increases significantly as the interaction strength approaches the critical value $U_c$, as well as when $\beta$ decreases. Consequently, if $L_y$ is initially chosen to be too small, the expected linear relationship between $\mathcal{F}(\Delta L_x)$ and $\Delta L_x$ is obscured. In practice, one must systematically increase $L_y$: if the mixed state is topologically nontrivial, the curve will progressively recover its linear form, at which point Eq.~\eqref{S_ls} remains valid.

\begin{figure}[htbp] 
    \centering
    \includegraphics[width=\linewidth]{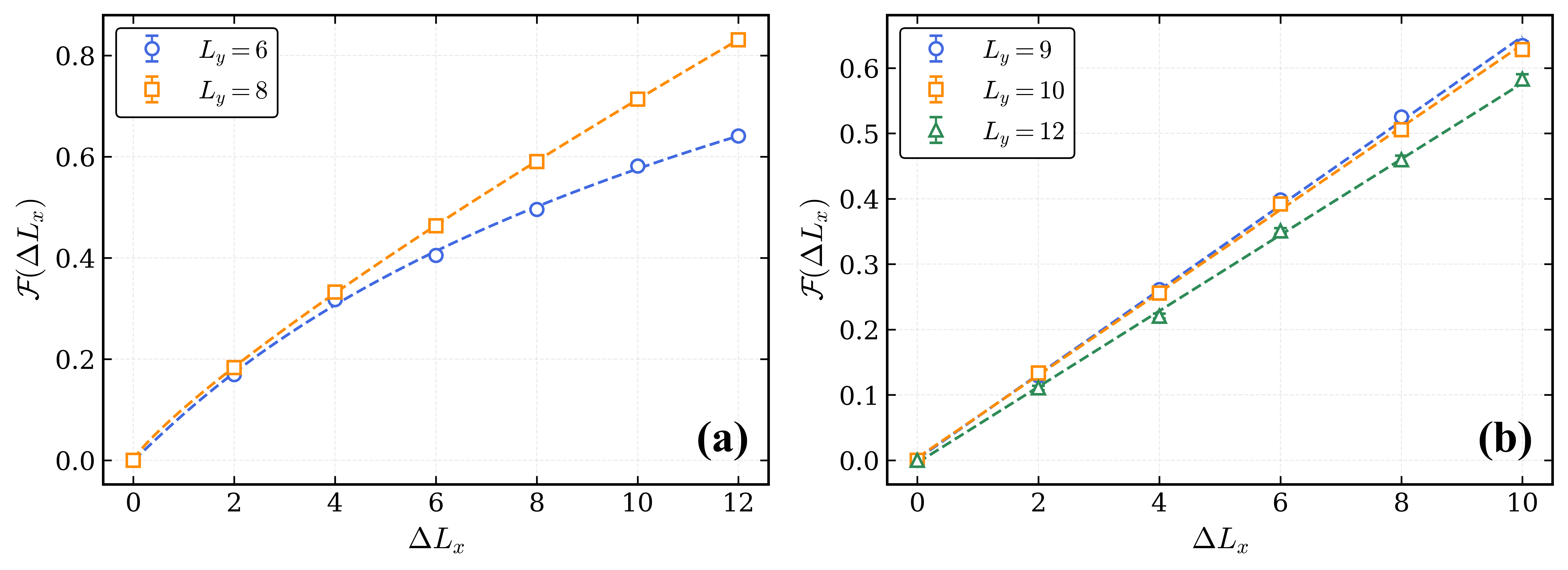}  
    \caption{
The calculations are performed at $\beta = 10.0$ and $U = 4.5$, which is close to but still below the critical point $U_c \approx 5.0$. 
(a) For small $L_y = 6$, the linear relation between $\mathcal{F}(\Delta L_x)$ and $\Delta L_x$ does not hold, whereas increasing to $L_y = 8$ causes the curve to bend upward and gradually recover a straight-line behavior (reference $L_x^0 = 7$). 
(b) Once $L_y$ becomes sufficiently large, Eq.~\eqref{S_ls} holds, as evidenced by the inverse dependence of the slope on $L_y$ (reference $L_x^0 = 9$).
    }
    \label{fig:S4}
\end{figure}

This behavior is demonstrated in Fig.~\ref{fig:S4}. For $\beta = 10.0$ and $U = 4.5$—a point within the topological phase but close to the critical transition $U_c = 5.0$—panel (a) shows that the linear relation breaks down for small $L_y$. However, as $L_y$ increases, the curve evolves, eventually converging to the expected linear dependence. Panel (b) confirms that once linearity is restored, Eq.~\eqref{S_ls} holds, evidenced by the observation that the slope decreases as $L_y$ increases.

\end{document}